**A Survey of Link Prediction Techniques**

Vivian Feng

Thomas Jefferson High School for Science and Technology

Computer Systems Research Lab

Mr. Kosek

May 1, 2023

# Table of Contents

**Abstract**

*The problem of link prediction, predicting if two nodes in a network have a connection between them, is a theoretical problem with numerous field-agnostic real-world applications. This paper investigates the efficacy of three classes of link prediction algorithms: local node similarity heuristics, the global index Random Walk with Restart, and Node2Vec embeddings. Furthermore, this paper provides insight into the performance of canonical link prediction algorithms on small graphs. The graphs included in this study are sampled from various domains, including infrastructure and ecological networks.*

## A Survey of Link Prediction Techniques

## Introduction

Graphs can represent structures like road networks, airplane flights, ecological communities, and social relationships. The task of link prediction, determining whether two unconnected nodes in a graph have an edge between them, is a theoretical problem with field-agnostic real-world applications. Because the structures of real-world networks vary greatly, no single link prediction algorithm is a panacea. Different link prediction algorithms have different efficacies. This study will investigate the efficacy of local similarity indices, random walk with restart global index, and Node2Vec embeddings on graphs not traditionally used in academic research.

The *Background* section provides a theoretical overview of the link prediction algorithms included in this survey. The *Development and Techniques* section discusses implementation and and *Discussion* provides an overview of results.

## Background

### I. Problem Definition

A graph is a collection of nodes linked together by edges. A link is equivalent to an edge. In this paper, the terms "vertex" and "node" will be used interchangeably. Likewise, the term "graph" and "network" will also be used interchangeably.

For undirected, unweighted graphs, the link prediction problem aims to find unobserved edges in an incomplete version — named *G'* — of the entire graph *G* (Refer to Figure 1 for a graphical representation of the problem). Link prediction has a multitude of practical use cases. For example, Facebook uses link prediction to suggest people users may know (Facebook, n.d.). Furthermore, cybersecurity professionals can identify potential lateral attacks in computer networks using node-embedding-based methods (Bowman, 2022).

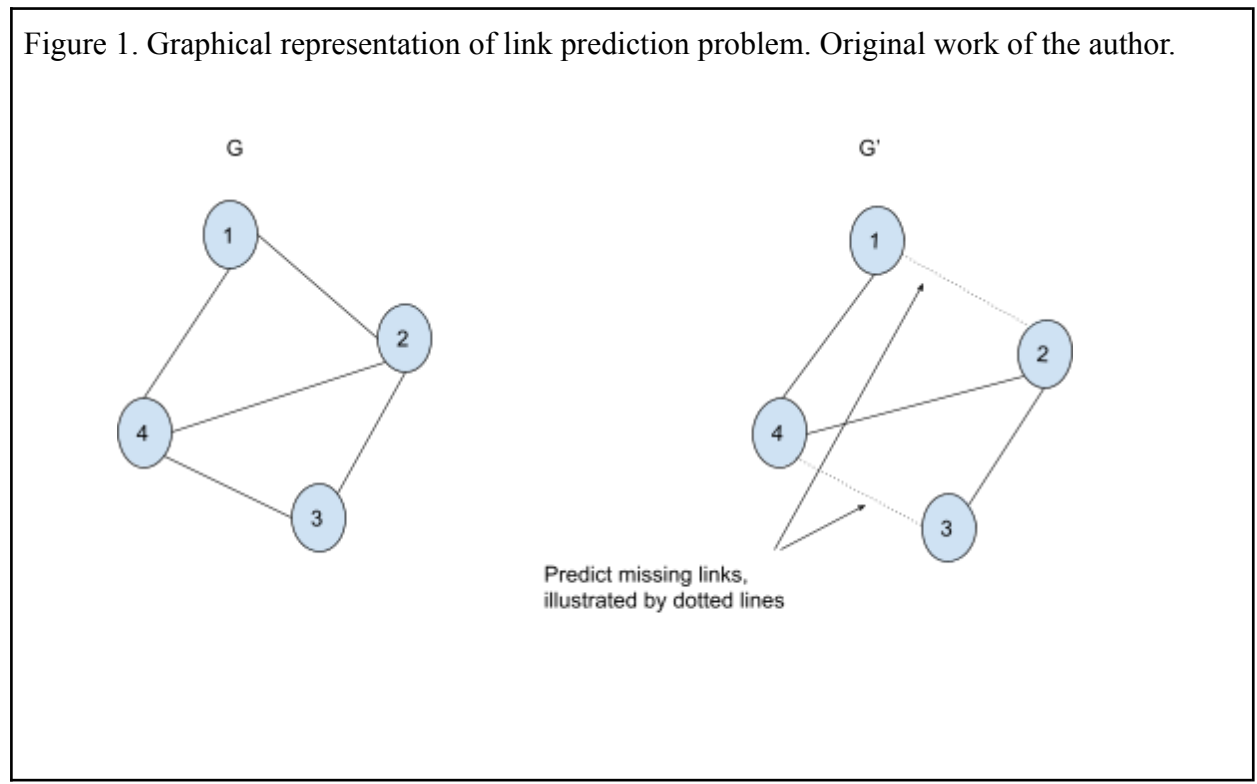

Figure 1. Graphical representation of link prediction problem. Original work of the author.

### *I.1 Evaluation metrics*

Link prediction algorithms assign pairs of nodes a numerical value which quantifies the likelihood that the pair of nodes has an edge between them. This numerical value, also known as a score, should be higher for pairs of nodes that are edges than for pairs of nodes that are not edges. Area under the Receiver Operating Curve (AUC) is the probability that a link prediction algorithm assigns an existing link a higher score than a nonexistent link. The higher the value of AUC, the better the link prediction algorithm being evaluated is. Algorithmically, this paper will estimate the AUC value using the formula (Lü & Zhou, 2011, p. 1152):

$$AUC = \frac{n' + 0.5n''}{n}$$

where out of *n* independent random selections of edges, *n'* is the number of times that an existing edge has a higher score than a nonexistent edge, and *n''* is the number of times that an existing edge has a score less than or equal to that of a nonexistent edge.

## II. Local Similarity Indices

Local similarity indices assign potential edges a score based on how many shared neighbors the two nodes composing the edge have. Local similarity indices apply a normalizing factor to the number of shared neighbors when computing the score of a potential edge. Although local similarity indices are not as accurate as global indices and machine-learning-derived methods, local similarity indices are fast, do not require extensive hyperparameter tuning, and are easily portable between graphs of different subject areas. The following sections detail some representative local similarity indices.

### II.1 Common Neighbors (CN)

Common Neighbors (CN) Index is denoted by $s_{cn}$. It is a general-purpose link prediction index, although its assumption that a greater number of neighbors corresponds to a greater probability of a link does not hold across all networks. For instance, mutual neighbors are better predictors of links in social networks than in protein interaction networks (Liu et al., 2023). The CN index is defined as such: given two nodes $x$ and $y$, where $N(x)$ denotes the set of node $x$'s neighbors, $|Q|$ denotes the number of elements contained in set Q, and $\cap$ denotes the intersection of two sets, the value of the Common Neighbors Index is (Lü & Zhou, 2011, p. 1153):

$$s_{cn} = |N(x) \cap N(y)|$$

### II.2 Hub-Promoted Index

The hub-promoted index assigns links that are closer to hubs (highly connected nodes that integrate the different components of a modular graph) with higher scores. This property is useful for quantifying the overlap in neighbors between nodes in metabolic networks (Ravasz et al., 2002). The Hub-Promoted Index ($s_{HP}$), where $k_x$ is the degree (number of neighbors of a node) of node $x$, is defined as (Lü & Zhou, 2011, p. 1154):

$$s_{HP} = \frac{|N(x) \cap N(y)|}{min(k_x, k_y)}.$$

### II.3 Hub-depressed Index

Rather than awarding hubs higher scores, the hub-depressed index penalizes hubs by dividing by the maximum degree between nodes $x$ and y. As such, nodes within the neighborhood of hubs have lower scores. The Hub-Depressed index is theoretical in nature and does not have many practical applications. The Hub-Depressed (HD) index is defined as (Lü & Zhou, 2011, p.1154):

$$s_{HD} = \frac{|N(x) \cap N(y)|}{max(k_x, k_y)}.$$

### II.4 Leicht-Holme-Newman index

Leicht et al. (2006) develop a model of vertex similarity in graphs based on the number of actual paths in a graph and the number of expected paths in the same graph with randomly connected vertices. This node similarity index is obtained by dividing the number of paths of length two in a graph by the expected number of paths of length two suggested by Leicht et al.'s model and dropping the multiplicative constants. The Leicht-Holme-Newman Index is defined as (Lü & Zhou, 2011, p.1154):

$$s_{LHN} = \frac{|N(x) \cap N(y)|}{k_x \times k_y} .$$

The denominator is thought to be proportional to the expected number of nodes and acts as a normalizing factor on the index.

### II.5 Adamic-Adar index

Adamic and Adar (2003) proposed this index to identify links in social networks. They built social networks by analyzing texts and hyperlinks on university student homepages and mailing lists. To place greater emphasis on more unique shared neighbors, Adamic and Adar use an inverse log scaling scheme. The Adamic-Adar Index is defined as (Lü & Zhou, 2011, p.1154):

$$s_{AA} = \sum_{i \in N(x) \cap N(y)} \frac{1}{log(k_i)} .$$

## III. Global Similarity Indices

Global similarity indices, such as the Random Walk with Restart (RWR) Index, assign a score to a candidate link by examining whole-graph connectivity. Global similarities are more computationally complex than local node similarity indices, but global indices tend to have higher accuracy. The RWR index is derived from the steady-state probability of a random surfer that is wandering from node to node, with probability $c$ that they will go to a neighboring node and probability $1$-$c$ that they return to the starting node. It is assumed that the surfer traversing the nodes will choose any neighbor with equal probability. The transition matrix is defined as P, where $P_{ij}$ is $1/k_i$ if nodes $i$ and $j$ have an edge, else $P_{ij} = 0$ (Lü & Zhou, 2011, p.1156). $P_{ij}$ refers to the element in matrix P at row $i$, column $j$. Nodes are identified with integers ranging between 0 and $N$-1, where $N$ is the number of nodes.

At steady state, the probability of where the surfer traversing the nodes will end up starting at node $x$ does not change. If $\mathbf{q_x}$ is this probability vector, and entry $y$ in $\mathbf{q_x}$ denotes the probability that the surfer will eventually end up at node $y$ starting at $x$, then (Lü & Zhou, 2011, p.1156):

$\mathbf{q_x}$ = cP$^T\mathbf{q_x}$ + (1 - c) $\mathbf{e_x}$  ($\mathbf{e_x}$ denotes the $x$-th column of the identity matrix which has the same dimensions as P).

Solving for $\mathbf{q_x}$ finds that  $\mathbf{q_x}$ = $(1 - c)(I - cP^T)^{-1}\mathbf{e_x}$.

The index is then defined as $q_{xy} + q_{yx}$, where $q_{xy}$ is $y$-th element of vector $\boldsymbol{q_x}$ (Lü & Zhou, 2011, p.1156).

**IV. Node2Vec**

In recent years, advancements in machine learning have led to better low-dimension representations of graphs in vector space. These vector representations are known as embeddings. A widely used embedding model is Node2Vec. The core component of the Node2Vec embedding model is the Skip-gram architecture. This architecture was originally developed for generating word embeddings in natural language processing applications, but Grover and Leskovec (2016) used it with random walks to generate graph embeddings. Embeddings, like those created by Node2Vec, are more computationally efficient than global similarity indices. The additional space required for the RWR index has the complexity of $O(V^2)$, where $V$ is the number of vertices in the graph. On the other hand, additional space requirements for embeddings are dependent on hyperparameters and do not scale quadratically with respect to graph size — a property that is very desirable when working with graphs that have millions of nodes.

*IV. 1 Explanation of Node2Vec Architecture*

The Skip-gram model generates an embedding for a corpus of words in lower-dimensional space by trying to predict which words fall within a given word's context window. This context window is the words next to the given word in a sentence. The embedding vector is the hidden layer of the one-layer neural network used to generate the probability vector that describes the probability of each word in the corpus following the given word (Vatsal, 2022). Figure 2 illustrates the location of embeddings in the Skip-gram architecture.

Figure 2. Location of embeddings. Image modified from Vatsal (2022).

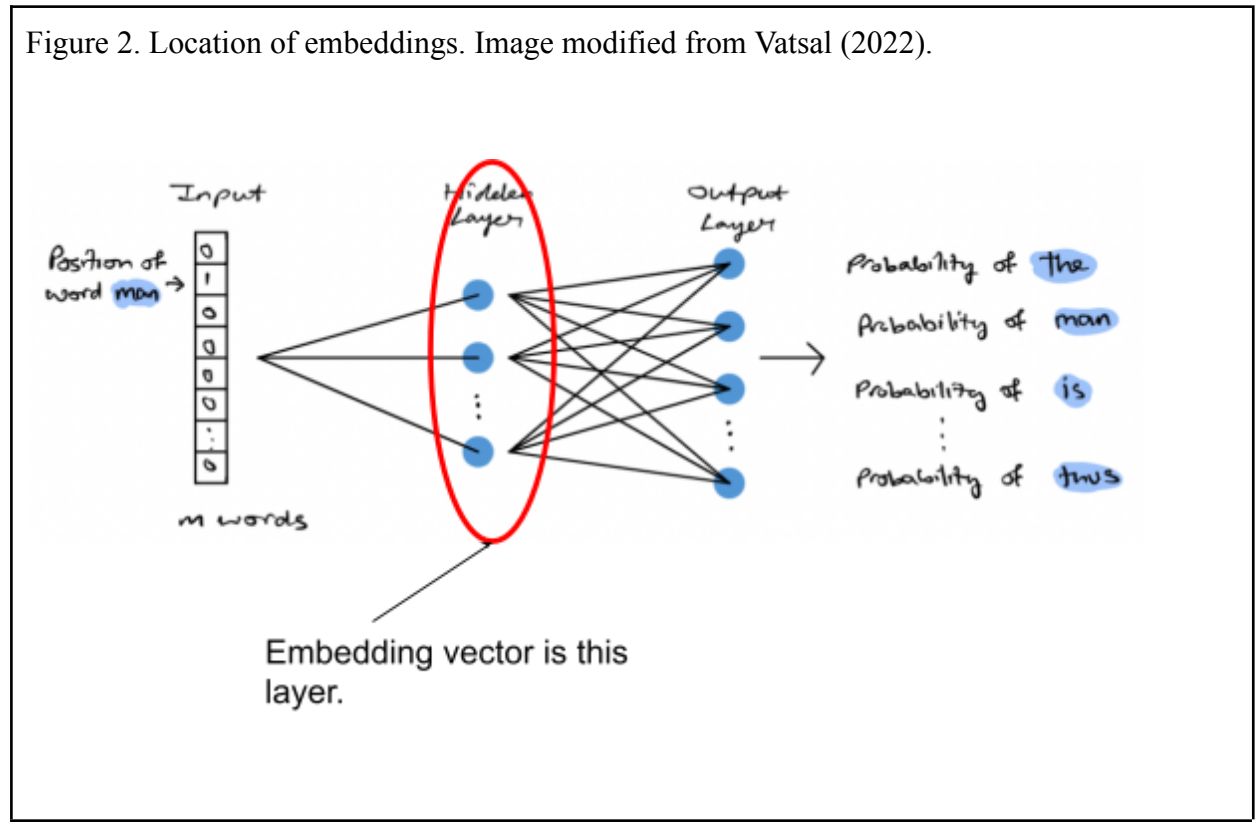

Analogously, the Skip-gram model can be applied to graphs. In the algorithmic framework Node2Vec, Grover and Leskovec (2016) developed a mapping of nodes to feature vectors in embedding space by applying the Skip-gram model to sequences of nodes sampled by weighted random walks of fixed lengths. Node2Vec treats random walks as "sentences" and nodes as "words" when inputting them into the Skip-gram model.

Using the random walks, the model learns the embedding vectors via stochastic gradient descent. Compared to other methods of node sequence sampling, such as depth-first search and breadth-first search, random walks have a lower big-O space and time complexity (Grover & Leskovec, 2016).

### IV.2 Visualization of Node2Vec Embeddings

Combined with dimensionality reduction algorithms such as Principal Component Analysis, the embeddings representing each node in a graph can be used to create visualizations that allow for easy identification of clusters of related nodes (Vatsal, 2022).

### IV.3 Generating Node Similarity Scores

In the embedding space, more similar nodes would have vectors that have less distance between them. Measures that determine the closeness of two vectors, such as cosine similarity and Euclidean distance, can therefore be used to generate scores for pairs of nodes. The embedding can also be used to train a logistic classifier.

## Development and Techniques

### Criteria for Success

The goal of this project is to implement and evaluate a selection of link prediction algorithms on small to medium-sized graphs sampled from a variety of subject areas.

### Overview

Using Python 3 and standard scientific Python libraries, I implemented commonly used local node similarity indices, the random walk with restart index, and Node2Vec embedding method. I developed programs in Visual Studio Code and used Microsoft Store's Python 3.10. Testing was done on a Lenovo IdeaPad Flex 5 with an 11th Gen Intel Core i7 Processor @ 2.80 GHz and 16.0 GB of RAM. The libraries used included BeautifulSoup (Richardson, 2022), Numpy (Harris et al., 2020), Scikit-learn (Pedregosa et al., 2011), Scipy (Virtanen et al., 2020), Gensim (Řehůřek, 2022), Pingouin (Vallat, 2018), and Matplotlib (Caswell et al., 2021). BeautifulSoup is used for web scraping; Numpy, Scikit-learn, and Scipy offer mathematical programing functionality; Gensim contains the skip-gram model necessary for implementing Node2Vec; Pingouin provides functions for statistical tests of significance; and Matplotlib offers graph visualization functionality.

### I.    Graphs Used for Testing

The five graphs being used for evaluation of link prediction algorithms are:

*eco-florida:* A network of different ecological communities in South Florida ecosystems (Rossi & Ahmed, *Eco-florida.zip,* n.d).

*inf-USAir97:* A network of US airline flights  (Rossi & Ahmed, *Inf-USAir97.zip,* n.d).

*LOTR:* A network of hyperlinks between different character pages on the Lord of the Rings fan wiki www.tolkiengateway.net. I obtained the graph using BeautifulSoup, a Python web-scraping library (Richardson, 2022).

*mammalia-dolphin-florida-social*: A network of interactions between dolphins in Cedar Key, Florida (Rossi & Ahmed, *Mammalia-dolphin-florida-social.zip,* n.d).

*road-chesapeake* - an infrastructure network (Rossi & Ahmed, *Road-chesapeake.zip,* n.d).

Refer to Table 1 for the characteristics of the graphs.

Table 1. Characteristics of graphs.

| Graph | Number of Nodes | Number of edges. | Average Degree |
|---|---|---|---|
| *eco-florida* | 128 | 2.1K | 32 |
| *inf-USAir97* | 332 | 2.1K | 12 |
| *LOTR* | 459 | 7.2K | 16 |
| *mammalia-dolphin-florida-social* | 151 | 1.6K | 20 |
| *road-chesapeake* | 30 | 170 | 8 |

## II.     Graph Class

The structure used for storing data for computations is a Graph class I coded in Python 3. In addition to convenience functions that find values commonly used in local node similarity indices, it also contains utility functions that return the adjacency matrix, adjacency list, and edge list representation of a graph. Undirected edges are represented as bidirectional directed edges. A Graph object can be initialized by a list of node pairs representing edges or read from an edge list stored in a file. A precondition is that the edge list must only contain space-separated pairs of nodes, and have no extraneous lines like comments or headers about graph size. The *Appendix* section titled "Graph Class" contains the code and comment-annotated functions. Nodes are represented with integers.

### III.    Sampling

Sampling is done by splitting the edge list of a graph into a training and test set using scikit-learn's train_test_split() function. For all graphs, 90% of the edges were used in the training set and 10% of the edges were used in the test set. Scikit-learn is a library that contains implementations of various data processing and machine learning algorithms (Pedregosa et al., 2011).

### IV.    Implementation of AUC

AUC is experimentally measured by $n$ random samplings. From the test set, an edge is randomly selected. Using the training set, its score, a numerical value that is a heuristic for the probability that a candidate is an edge, is computed and compared to the score of a random nonexistent edge. The times the score of a pair of nodes forming an edge is greater than a nonexistent edge ($n'$) and the times the score is less than or equal to the nonexistent edge ($n''$) are recorded. The AUC score is then calculated according to the formula given in section *I.1 Evaluation Metrics* of *Background*. Refer to the *Appendix* for its implementation under the title "AUC function".

## V.     Implementation of the Random Walk with Restart Index and Local Similarity Indices

The local similarity algorithms were implemented using the Python standard library functions, according to the formulas described in the Background section. Experiments included a novel variation of the LHN index. The normalizing factor in the denominator takes the log of the product of the degree of each node, and is defined as such: $s = \frac{|N(x) \cap N(y)|}{log(k_x \times k_y)}$.

To avoid division-by-zero errors, the value of a quotient is assumed to be zero if there is a zero denominator in the variation of the LHN index and the Adamic-Adar index. The implementations of the similarity indices are found in the *Appendix* under "NodeSimilarities.py".

The Random Walk with Restart Index is implemented following the formula given in the *Background* using NumPy, a linear algebra library (Harris et al., 2020). Implementation-wise, I compute $M = (1 - c)(1 - cP^T)^{-1}$ instead of $\mathbf{q_x}$ to find the value of the Random Walk with Restart Index. If $M = (1 - c)(1 - cP^T)^{-1}$, then the sum $M_{xy}$ and $M_{yx}$ equals $q_{xy} + q_{yx}$, because multiplication by $\mathbf{e_x}$ gets the $x$-th column of the matrix $M$.

## VI.    Node2Vec Embedding Generation

The embedding generation pipeline was based on the one provided by Grover and Leskovec (2016). Generation is divided into three main parts: reading the graph, obtaining walks that serve as the "sentences" for the Skip-gram model, and generating embeddings that correspond to each node. As described in the *Appendix* under "randwalkgraph.py", the function getWalks() returns a list of random walks with length $l$, and $r$ walks per node that use the node as a starting point. If the argument *weighted* is True, then the random walk used to sample nodes is a weighted random walk with alias sampling. Hyperparameters $p$ and $q$ control the probability weightings of the current node's neighbors.

Let the current node be $x$. The unnormalized probability distribution for choosing the next node in the walk is defined with respect to the distance of node $x$'s neighbors from the node which comes before $x$ in the walk. Let $t$ be the node that was traversed before $x$ in the walk. Table 2 contains a description of the unnormalized transition probability weightings.

Table 2. Transition probabilities.

| Unnormalized Transition Probabilities | |
| --- | --- |
| 1 | Node is a neighbor of $t$ and a neighbor of $x$ |
| $1/p$ | Node is $t$ |
| $1/q$ | Node is a neighbor of $x$ but not a neighbor of $t$ |

If *weighted* is False, then a random walk with restart is utilized, and the probability that the node will stay at its current node, controlled by the hyperparameter $c$, is used. To the best of my knowledge, this is a novel variation of Node2Vec embedding generation. The function learnFeatures() creates the embeddings and calls getWalks(). It passes the relevant parameters to getWalks(), and uses the returned walks to train the Word2Vec model from Gensim (Řehůřek, 2022). The final embeddings are then saved to a file.

For alias sampling to be practical, the selection of the next node needs to take place in constant time (Grover & Leskovec, 2016). To accomplish constant time sampling, Vose's alias sampling method is used. Figure 3 provides a description of the aforementioned algorithm.

Figure 3. Description of Vose's alias sampling algorithm. Adapted from Schwarz (2011).

```python
def alias_setup(self,p,q):
    # maps each edge to the appropriate probability and edge samples
    self.aliasTable = dict()
    # for each edge
        # initialize alias and prob
        # create two worklists
        # scale each one by sum(prob) and n
        # for each scaled prob:
            # if les than one: add i to small
            # else add to large
        # while !small and !large empty
            # pop from small, call it l
            # pop from large, call it g
        # set prob[l] to p[l]
        # set alias to g
        # set pg = (pg+pl)-1
        # if pg < 1 add g to small
        # otherwise add g to large
        # while large is not empty
            # remove the first element from large, call it g
            # set prob[g] = 1
        # while small is not empty
            # remove first element from small call it l
            # set prob[l] = 1
        # sampling
        # filp a fair n sided die
        # flip the coin that comes up heads with prob[i]
        # if heads return i
        # else return alias[i]
```

**VII.**     **Node2Vec Prediction Implementation**

I used methods outlined in subsection **VI** in *Development and Techniques* to create node embeddings for a graph built from the set of edges in the training set. The procedure for training a logistic regression classifier for use in link prediction, similar to the method proposed by Grover and Leskovec (2016), is as follows:

1. Given two nodes $u$ and $v$, combine them using an operator, such as pairwise multiplication.

2. The resultant vector is then used as the input into the logistic regression model. The output is a value between 0 and 1, determining the probability of the pair of nodes having an edge (1) or having no edge (0).

The model I use is Scikit-learn's logistic regression model (Pedregosa et al., 2011). The training data set of the model contains an even distribution of positive and negative examples. The positive examples are the entirety of the edges not withheld for testing. The negative examples are an equal number of randomly generated nonexistent edges. The value generated by the logistic regression classifier is treated as the score of a potential edge and used to compute AUC.

**VIII. Visualization**

To enhance the interactivity of the project, I created an interface that allows a user to generate different Node2Vec embedding representations. After a high dimensional embedding for a graph is created, they are projected into the 2D plane using Principal Component Analysis followed by T-distributed Stochastic Neighbor Embedding, two dimension reduction projection algorithms. Figure 4 contains a screenshot of the user interface. The parameters in Figure 4 represent the following:

- – d = dimension of embeddings

- – r = number of walks per node

- – l = length of walks

- – k = context window size (number of nodes before and after a given node to consider when generating embeddings)

- – p, q = weights controlling the alias-sampled walk

Figure 4. Image of user interface.

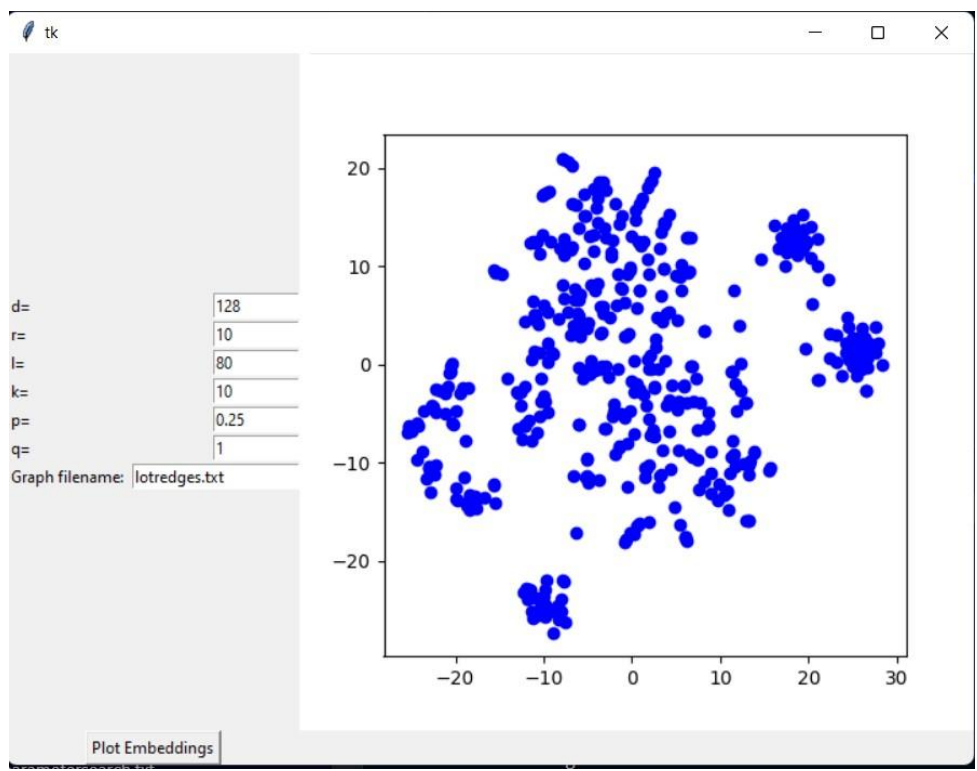

**VIIII. Data Collection and Testing of Algorithms**

Initially, graphs (see **I. Graphs used for Testing** in *Development and Techniques*) were stored in a text file containing the node pairs separated by new lines. Each graph was loaded into memory by reading from the file and storing it in a Graph object. The edges were then split between a training set and a test set in a 90-10 split. Next, an incomplete graph was constructed using edges in the training set. The incomplete graph was then used to evaluate the AUC. I examined the effect of different hyperparameter values of $c$ on AUC for the Random Walk with Restart index, performance differences between local node similarity indices, the effect of embedding dimension on the performance of Node2Vec embeddings created with alias-sampled random walks, and the effect of $c$ for Node2Vec embeddings created using random walks with restart. Testing was done with the AUC function described earlier. Across different independent variables, I used different partitions of training and test sets. For different levels of the same independent variable, however, I used the same partition for training and test set data — a repeated-measures design. Each variable was tested on 100 different partitions and results were analyzed using a repeated-measures ANOVA.

No formal proof of correctness was performed on my implementations. However, AUC values obtained for the graph *Inf-USAir97* were consistent with the values obtained by Lü & Zhou (2011) for the same graph (referred to as USAir by Lü & Zhou). The values for other graphs were also in the neighborhood of Lü and Zhou's results. Node2Vec results were compared with the results from the reference implementation supplied by Grover and Leskovec (2016), and AUC for link prediction using embeddings created with the reference implementation are the same as scores obtained using embeddings created by my own implementation.

**X. Limitations**

The number of algorithms included and hyperparameters in this survey was limited by the amount of time available to implement them. Furthermore, the testing code was rewritten multiple times to improve experiment design, so data collection took longer than expected.

**Discussion**

Charts in each section depict network-specific experiment results.

**I.   Effect of Different Hyperparameter Values of $c$ on AUC for the Random Walk with Restart Index Performance**

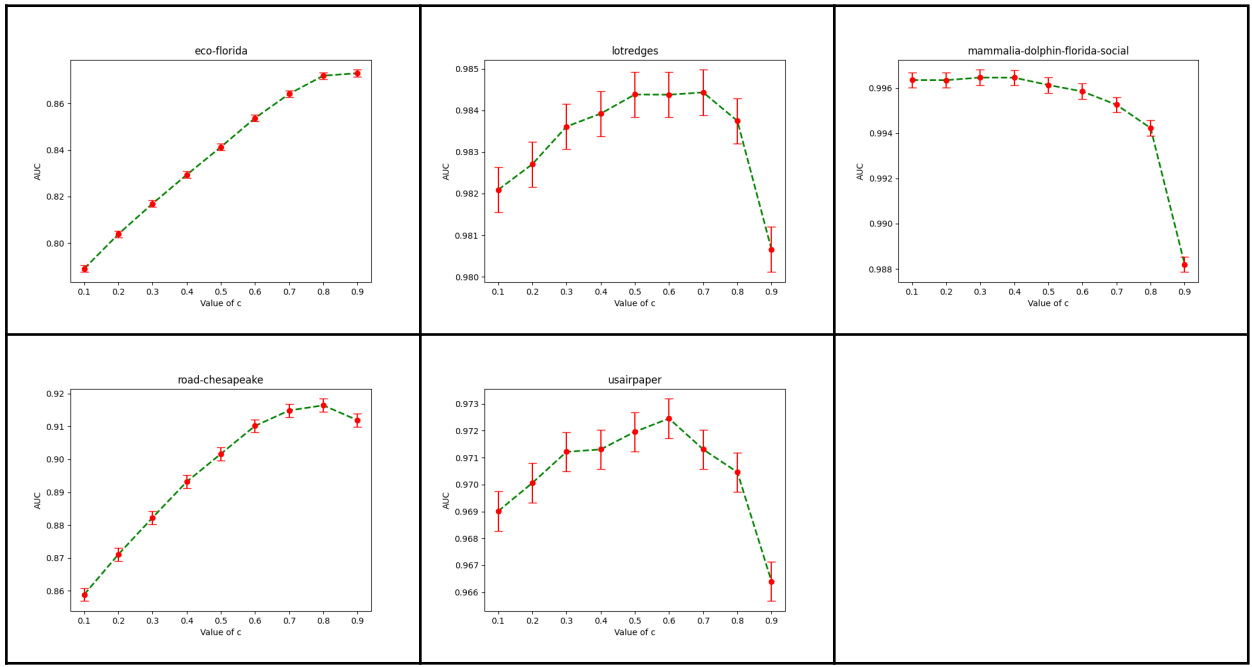

Except for the graph *mammalia-dolphin-florida-social*, the accuracy of the Random Walk with Restart index, as measured by AUC, increases as $c$ increases, then tapers off. Values of $c$ less than 0.5 appear to produce worse results than values greater than 0.5. *usairpaper* is another name for *Inf-USAir97*. The data, in general, is normally distributed, and sphericity violations were accounted for in the repeated-measures ANOVA with the Greenhouse-Geisser correction. Error bars show the 95% confidence interval, wherein the standard error is calculated using the formula $\sqrt{\frac{MS_{error}}{n}}$ (Loftus & Mason, 1994). $MS_{error}$ is the within-subject mean squared error for the repeated-measures ANOVA and $n$ is the number of samples. Statistically significant differences exist between the different levels of *c* for all tested graphs at the 0.05 significance level.

**II.    Differences Between Local Node Similarity Indices**

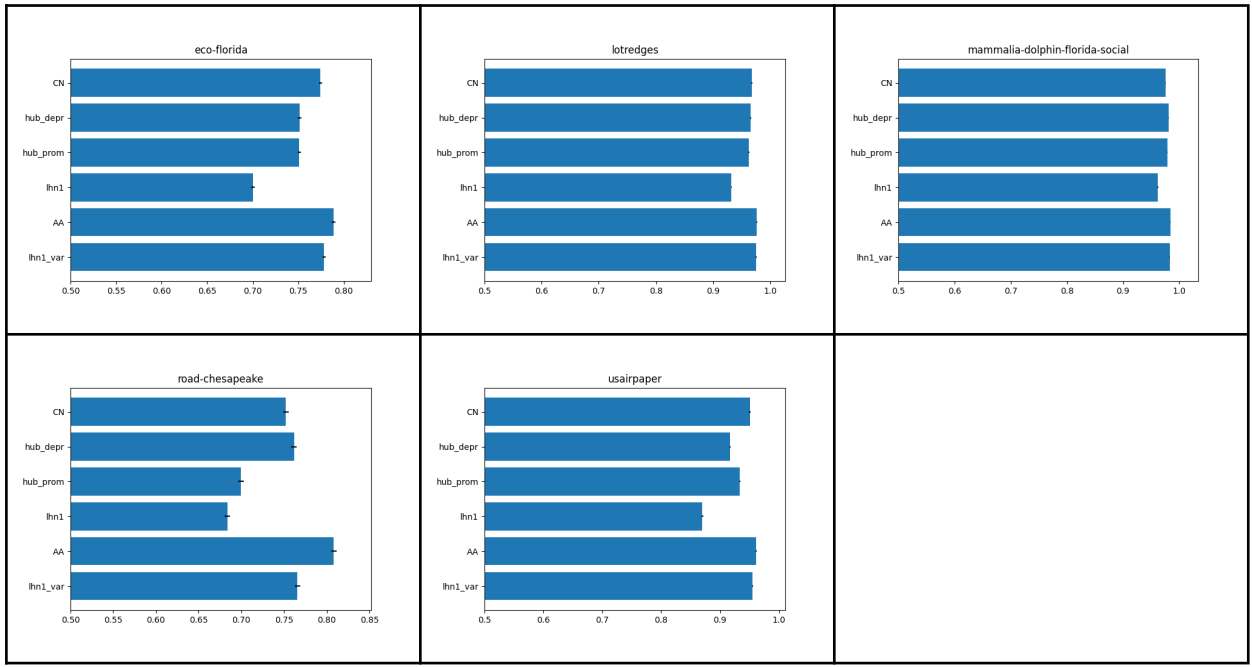

Across all five graphs, the Adamic-Adar index has the best performance (in terms of AUC), whereas the Leicht-Holme-Newman (lhn1) index has the worst performance. *usairpaper* is another name for *Inf-USAir97*. The key for labels in the charts above are as follows:

- CN = Common Neighbors

- hub_depr = Hub Depressed Index

- hub_prom = Hub promoted index

- lhn1 = Leicht-Holme-Newman

- AA = Adamic Adar

- lhn1_var = Leicht-Holme-Newman variation

The index lhn1_var, as proposed in *Development and Techniques* section **V**, appears to perform as well or better than common neighbors, hub-promoted, and hub-depressed index across the five graphs. The AUC values obtained for each local node similarity index were, for the most part, normally distributed. For all the different graphs, the repeated-measures ANOVA found statistically significant differences between the performance of each index at the 0.05 level. Violations of sphericity were accounted for with the Greenhouse-Geisser correction, and both the uncorrected and corrected p values were less than 0.05. Error bars were calculated with the same methodology as described in *Discussion* section **I**.

**III.    Effect of Embedding Dimension on Performance of Node2Vec Embeddings Created with Alias-sampled Random Walks**

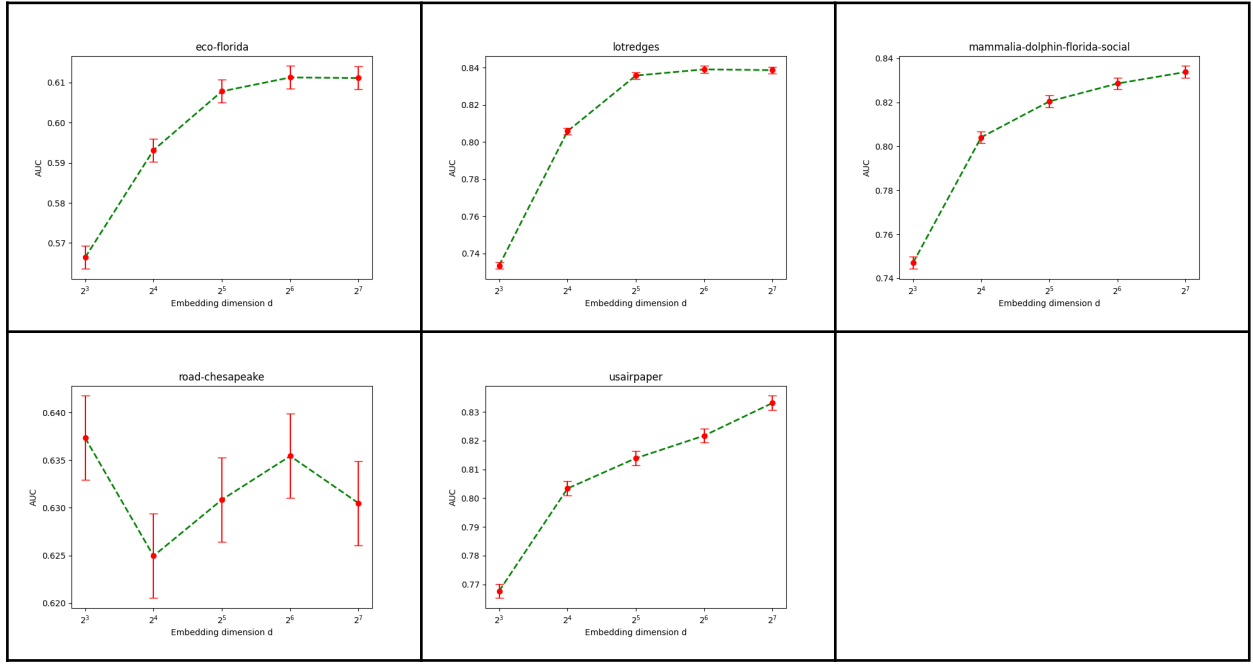

For all graphs except for *road-chesapeake*, as embedding dimension increases, the performance of logistic regression trained on alias-sampled Node2Vec embeddings increases. The reported AUC values here for Inf-USAir97 (labeled here as *usairpaper*) are slightly less than those reported by Kumar et al. (2020) for the same airline graph, but these discrepancies and other AUC discrepancies found during testing are most likely due to differences in parameter choice, rather than systemic flaws.

Different values of p and q were selected for each graph, but parameters other than d — context window size, length of walk, number of walks per node starting at that node, and epochs used to train the Skip-gram model — were held constant across all graphs. *Road-chesapeake* likely does not exhibit the pattern of increasing AUC with dimensionality because the graph has 170 edges, which is about ten times less than the number of edges the other graphs have. At higher dimensions, the issue of overfitting would lead to worse performance. The data collected for each level of dimensionality was normally distributed, and violations of sphericity were accounted for with the Greenhouse-Geisser correction. A repeated-measures ANOVA found statistically significant differences between different dimensions for all graphs at the 0.05 significance level. Error bars were calculated with the same methodology as described in *Discussion* section **I**.

**IV.     Effect of *c* for Node2Vec Embeddings Created with Random Walks with Restart**

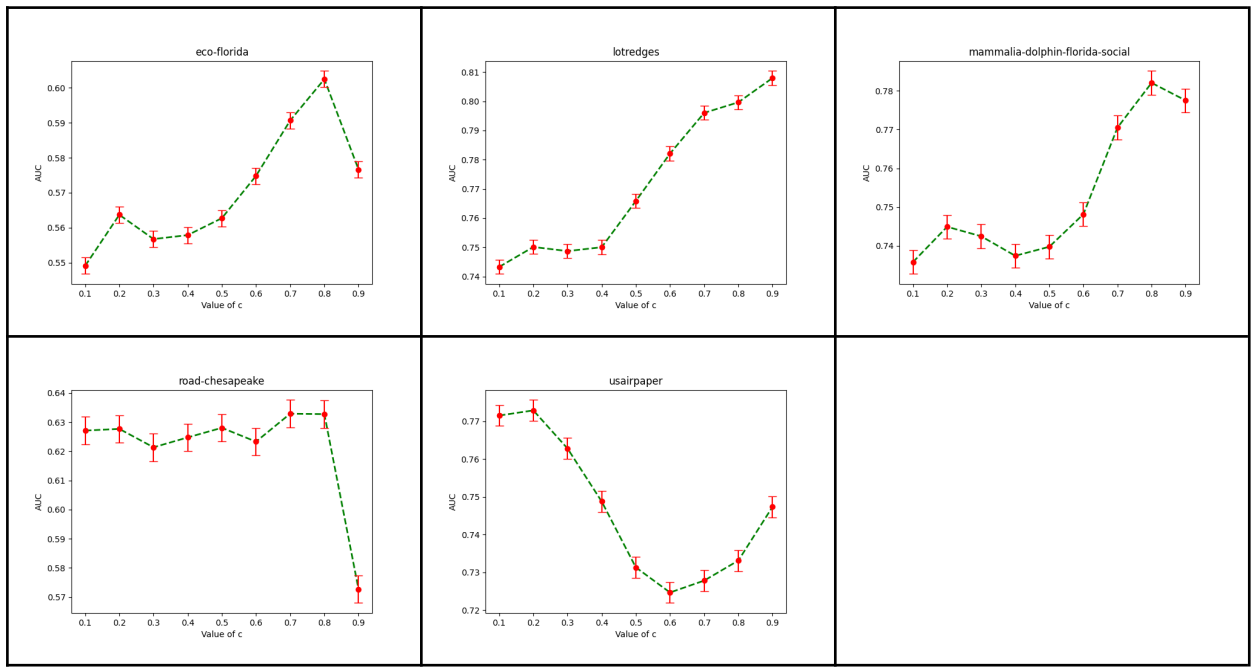

For *eco-florida*, *lotredges*, and *mammalia-dolphin-florida-social* graphs, as the value of *c* increases, AUC increases. However, *road-chesapeake* and *Inf-USAir97* (labeled "usairpaper") exhibit declining AUC with respect to *c*. As such, there is no obvious trend between AUC and *c*. Overall, the AUC for random walk with restart generated embeddings appear slightly less than alias-sampled embeddings, but a thorough hyperparameter search may lead to different results.

The values for the hyperparameters context window size, length of a walk, number of walks per node starting at that node, and epochs used to train the Skip-gram model were the same as the ones used to create alias-sampled random walk-generated embeddings in section **III** of *Discussion*. The dimension of embeddings for the random walk with restart embeddings was $d = 128$ for all graphs.

The data collected for each level of $c$ was generally normally distributed, and the Greenhouse-Geisser correction was used to account for violations of sphericity. A repeated-measures ANOVA reported statistically significant differences between different levels of $c$ for all graphs at the 0.05 significance level. Error bars were calculated with the same methodology as described in *Discussion* section **I**.

## Further Recommendations

As noted elsewhere, a more thorough hyperparameter search could yield better performance for the algorithms explored in this survey. While this survey did not examine runtime, the implementations of the Graph class and each of the algorithms could be optimized, and perhaps turned into a link prediction Python library, in the future.

**Appendix**

**Graph Class**

```python
import numpy
from sklearn.model_selection import train_test_split
import random
import pickle
class Graph:
    """
    Graph object contains a graph. On initialization, it initializes an
empty adjacency list.
    Graph is undirected.
    """
    def __init__(self):
        self.adjList = dict()
        self.edgePrs = []
        self.nodeList = []
        # self.size = -1 # Size = num Nodes
        self.revlist = dict()

    def readFromFile(A,filename):
        """
        Builds graph from file containing edge pairs
        """
        with open(filename,'r') as f:
            lines = f.readlines()
            for l in lines:
                l = l.split()
                start = int(l[0])
                target = int(l[1])
                A.addEdge(start,target)
                # A.edgePrs.append((start,target))
        A.nodeList = list(A.adjList.keys())
        A.revlist = dict()
        for k in range(len(A.nodeList)):
            A.revlist[A.nodeList[k]] = k
        # print("revlist",A.revlist)
        # print("edges",A.edgePrs)

    def readFromEdgeList(A,l):
        """
```

```python
        Builds from list containing tuples of edges between nodes
        """
        for u,v in l:
            A.addEdge(u,v)
        A.nodeList = list(A.adjList.keys())
        A.revlist = dict()
        for k in range(len(A.nodeList)):
            A.revlist[A.nodeList[k]] = k

    def toAdjMat(self, myrev = None):
        if myrev is None:
            myrev = self.revlist
        # mat = numpy.zeros(shape = (self.size,self.size))
        # use the indices of the nodelist to assign things
        # 1. reverse nodelist

        # myMat = numpy.zeros((self.size,self.size))
        # for u,v in self.edgePrs:
        #     i1 = revlist[u]
        #     i2 = revlist[v]
        #     myMat[i1,i2] = 1
        #     myMat[i2,i1] = 1
        # return myMat

        # 2. create adjmat with length of nodelist
        arr = numpy.zeros(shape=(len(myrev),len(myrev)))
        # print("edgepairs: ",self.edgePrs)
        for u,v in self.edgePrs:
            # print(u,v)
            n1 = myrev[u]
            n2 = myrev[v]
            arr[n1,n2] = 1
            arr[n2,n1] = 1
        return arr
    def addEdge(A,source,target):
        if source not in A.adjList:
            A.adjList[source] = set()

        if target not in A.adjList:
            A.adjList[target] = set()
        if target not in A.adjList[source]:
            A.adjList[source].add(target)
            A.adjList[target].add(source)
            A.edgePrs.append((source,target))
            # A.edgePrs.append((target,source))
```

```python
    # A.size = max(A.size,max(source,target))

def getDegree(A,node):
    """
    Returns the degree of a node
    """
    return len(A.adjList[node])

def getSharedNeighbors(A,u,v):
    """
    Returns the shared neighbors of two nodes
    """
    shared = A.adjList[u].intersection(A.adjList[v])
    # for node in A.adjList[u]:
    #     if node in A.adjList[v]:
    #         shared.append(node)
    return shared
def getNumberSharedNeighbors(A,u,v):
    """
    Returns the total number of shared neigbors
    """
    return len(A.getSharedNeighbors(u,v))

def testProbeSplit(A,test_size=0.1):
    """
    Divides the edges into training and probe sets. 90 10 split
    """
    # randomly sample 90% of edges into a train and probe set
    train,probe = train_test_split(A.edgePrs,test_size=test_size)
    # trainsize = len(train)
    # for i in range(trainsize):
    #     u,v = train[i]
    #     train.append((v,u))

    # testsize = len(probe)
    # for i in range(testsize):
    #     u,v = probe[i]
    #     train.append((v,u))
    return train,probe
# def pageRank(self,transitionMatrix,n1,n2):
#     i1 = self.revlist[n1]
#     i2 = self.revlist[n2]
#     return transitionMatrix[i1,i2] +
def toFile(self,name):
```

```python
        with open(name,'w') as f:
            for u,v in self.edgePrs:
                print(u,v,file=f)
                print(v,u,file=f)
            f.close()
    def getNeigbors(A,node):
        return A.adjList[node]

    def nonexistentLink(A,node):
        randval = random.choice(A.nodeList)
        # print(A.nodeList)
        while randval in A.adjList[node]:
            randval = random.choice(A.nodeList)
            # print("switched")
        return randval
```

**randwalkgraph.py**

```python
Python
import graph
import numpy as np
import random
from typing import *
class RandWalkGraph(graph.Graph):
    def __init__(self):
    super().__init__()

    def alias_setup(self,p,q):
    # maps each edge to the appropriate probability and edge samples
    self.aliasTable = dict()
    # for each edge
        # initialize alias and prob
        # create two worklists
        # scale each one by sum(prob) and n
        # for each scaled prob:
        # if les than one: add i to small
        # else add to large
        # while !small and !large empty
        # pop from small, call it l
        # pop from large, call it g
        # set prob[l] to p[l]
        # set alias to g
        # set pg = (pg+pl)-1
        # if pg < 1 add g to small
        # otherwise add g to large
        # while large is not empty
        # remove the first element from large, call it g
        # set prob[g] = 1
        # while small is not empty
        # remove first element from small call it l
        # set prob[l] = 1
        # sampling
        # filp a fair n sided die
        # flip the coin that comes up heads with prob[i]
        # if heads return i
        # else return alias[i]
    for u,v in self.edgePrs:
        for pr in [(u,v),(v,u)]:
        prev = pr[0]
        curr = pr[1]
        vals = self.getNeigbors(curr)
```

```python
prevSet = self.getNeigbors(prev)
N = len(vals)
probVec = np.zeros((N,))
idx = 0
idxToNode = dict()
prob = np.zeros((N,))
alias = np.zeros(N)
for v in vals:
        if v == prev:
                pi = 1/p
        elif v in prevSet:
                pi= 1
        else:
                pi = 1/q
        idxToNode[idx] = v
        probVec[idx] = pi
        idx += 1
# probVec = np.array(unnormedProbs)
probVec/=np.sum(probVec)
scaledProbs = N*probVec
# randomly sample an index from this set of nodes
small = []
large = []
for i in range(scaledProbs.shape[0]):
        if scaledProbs[i] < 1:
                small.append(i)
        else:
                large.append(i)
while len(small) > 0 and len(large) > 0:
        l = small.pop()
        g = large.pop()
        prob[l] = scaledProbs[l]
        alias[l] = g
        scaledProbs[g] = (scaledProbs[l]+scaledProbs[g]) - 1
        if scaledProbs[g] < 1:
                small.append(g)
        else:
                large.append(g)
while len(large) > 0:
        g = large.pop()
        prob[g] = 1
while len(small) > 0:
        l = small.pop()
        prob[l] = 1
```

```python
            self.aliasTable[pr] = dict(prob = prob,alias =
alias,idxToNode=idxToNode)

            # choiceIdx = np.random.choice(probVec.shape[0],p=probVec)
       # print(self.aliasTable)

       def aliasSample(self,neighborSet,startnode, prevNode,p,q):
       """
       Now with Linear Time sampling!!
       https://www.keithschwarz.com/darts-dice-coins/
       1/p if dtx = 0
       1 if dtx = 1
       1/q if dtx = 2
       """

       if prevNode < 0:
            return random.choice(list(self.getNeigbors(startnode)))
       # print(prevNode,startnode)
       if (prevNode,startnode) in self.aliasTable:
            alias = self.aliasTable[prevNode,startnode]["alias"]
            probs = self.aliasTable[prevNode,startnode]["prob"]
            idxtoNode = self.aliasTable[prevNode,startnode]["idxToNode"]
            choice = np.random.randint(alias.shape[0])
            # print(choice)
            r = np.random.random()
            if r < probs[choice]:
            return idxtoNode[choice]
            else:
            return idxtoNode[alias[choice]]

       def weightedWalk(self,x,length,p,q):
       """
       Call alias setup beforehand
       Returns a sequence of random walks starting at node x, weighted by
the parameters p and q
       x = starting node
       length = length of random walk
       p,q = parameters to weight graph
       """
       walk = [x]
       for i in range(length):
            curr = walk[-1]
```

```python
            possNext = self.getNeigbors(curr)
            prev = -1
            if len(walk) >= 2:
            prev = walk[-2]
            s = self.aliasSample(possNext,curr,prev,p,q)
            walk.append(s)
        return walk
        def randomWalkwithRestart(self,x:int,length:int,c:float=0.9) ->
List[int]:
        """
        Implements a random walk with restart, in contrast to the Weighted
Random Walk used by the original Node2Vec paper
        Probability c that surfer goes to next node, probability 1-c that
surfer returns to start node. Returns a list containing the walks
        """
        walk = [x]
        for i in range(length):
            curr = walk[-1]
            possNext = list(self.getNeigbors(curr))
            r = np.random.rand()
            if r < c:
            walk.append(random.choice(possNext))
            else:
            walk.append(x)
        return walk

def getWalks(graph:"RandWalkGraph",l,r, p,q, c=.1,weighted=True):
        walks = []
        for i in range(r):
        print("Walk ",i," obtained")
        if weighted:
            print("Using alias sampling")
        else:
            print("using random walk with restart")
        for u in graph.nodeList:
            if weighted:
            walk = graph.weightedWalk(u,l,p,q)
            else:

            walk = graph.randomWalkwithRestart(u,l,c)
            # print("Walk: ",walk)
            walks.append(walk)
        return walks
def learnFeatures(graph:"RandWalkGraph",d,r,l,k,p=1,q=1,c=0.1,filename =
"embeddings.eb",weighted=True):
```

```python
    """
    graph = G
    Dimensions = d
    Walks per Node = r
    Length of walks = l
    Context size = k
    Parameters p and q
    """
    walks = getWalks(graph,l,r,p,q,c=c,weighted=weighted)
    print("Training model")
    print(len(graph.edgePrs))
    # print(len(walks))
    model =
Word2Vec(sentences=walks,vector_size=d,window=k,min_count=1,workers=4,ep
ochs=10)
    print("Model done training!")
    model.wv.save_word2vec_format(filename)

def readFromFeatures(filename):
    """
    Takes embedding file as input
    Readsd from that embedding file
    """
    nodeToEmbedding = dict()
    with open(filename,"r") as f:
    lines = f.readlines()
    n, emb_size = tuple(map(int,lines[0].split()))
    all_embeddings = []
    for line in lines[1:]:
        # line = lines[i]
        # vals = line.split()
        # myId = int(vals[0])
        # all_embeddings.append(list(map(float,vals[1:])))
        # idtoIdx[myId] = len(all_embeddings)-1
        line = line.split()
        node = int(line[0])
        emd = np.array(list(map(float,line[1:])))
        nodeToEmbedding[node] = emd
    return nodeToEmbedding
```

**NodeSimilarities.py**

```python
from typing import *
from graph import *
from math import log10,sqrt
#mygraph: "Graph"= pickle.load('savedLotr.pkl')
def commonNeighborsimilarity(g:"Graph",node1:int,node2:int) -> int:
    """
    Returns a similarity value given two nodes
    """
    return g.getNumberSharedNeighbors(node1,node2)

def hubDepressedsimilarity(g:"Graph",node1:int,node2:int) -> float:
    """
    Return the hub depressed
    """
    sharedneighbors = g.getNumberSharedNeighbors(node1,node2)
    return sharedneighbors/min(g.getDegree(node1),g.getDegree(node2))

def hubpromottedsimilarity(g:"Graph", node1:int, node2:int) -> float:
    sharedneighbors = g.getNumberSharedNeighbors(node1,node2)
    return sharedneighbors/max(g.getDegree(node1),g.getDegree(node2))

def  lhn1similarity(g:"Graph",node1:int,node2:int):
    shared = g.getNumberSharedNeighbors(node1,node2)
    return shared/(g.getDegree(node1) * g.getDegree(node2))

def adamicadarSimilarity(g:"Graph",node1:int,node2:int) -> float:
    """
    Adamic-Adar Index: sum of 1/log(k) of shared neighbors
    """
    shared = g.getSharedNeighbors(node1,node2)
    toRet = 0
    for k in shared:
        if(g.getDegree(k)!=1):
            toRet += 1/(log10(g.getDegree(k)))
    return toRet

def mySimilarityIndex(g:"Graph",node1:int, node2:int):
    """
    Try using size of of shared neighbors as a coefficient to the
adamic-adar index
    numbershared/(log product of node1 and node1)
    """
```

```
val = log10(g.getDegree(node1))+log10(g.getDegree(node2))
if abs(val) < 1e-9:
    return 0
return g.getNumberSharedNeighbors(node1,node2) / val
```

**AUC function**

```Python
def AUC(g:"Graph",train,test,simFunc,n=1000) -> float:
    """
    AUC = (n'+0.5n")/n, take n to be the size of the test set
    : Provided the rank of all non-observed links, the AUC value can
be interpreted as the probability that a randomly
chosen missing link (i.e., a link in E
P
) is given a higher score than a randomly chosen nonexistent link (i.e.,
a link in U -E). In
the algorithmic implementation, we usually calculate the score of each
non-observed link instead of giving the ordered list
since the latter task is more time consuming.4
Then, at each time we randomly pick a missing link and a nonexistent
link to
compare their scores, if among n independent comparisons, there are n
'
times the missing link having a higher score and n
''
times they have the same score, the AUC value is

    """

    nprime = 0
    n2prime = 0
    # n = 0
    traingraph = Graph()
    traingraph.readFromEdgeList(train)

    for i in range(n):
        # choosing existing link
        u,v = random.choice(test)
        if u not in traingraph.adjList or v not in
traingraph.adjList:
            probescore = 0
        else:
            probescore = simFunc(traingraph,u,v)
        # choosing nonexistent link
        """"Depending on algorithm being evaluated, the nonexistent
edge sampling code may change. This AUC function is for evaluating
local similarity indices. However, the core idea is the same:
choose two random nodes, and while the pair forms and existing
edge, choose another random node to get a nonexistent edge"""
```

```python
            rand = random.choice(traingraph.nodeList)
            other = traingraph.nonexistentLink(rand)
            nonexistent = simFunc(traingraph,rand,other)
            if probescore > nonexistent:
                    nprime += 1
            else:
                    n2prime += 1
    return (nprime+n2prime/2)/n
```